\documentclass[sigconf,natbib=true,anonymous=false]{acmart}

\usepackage{url}
\usepackage{xcolor}
\usepackage[ruled]{algorithm2e}
\usepackage{multirow}
\usepackage{verbatim}
\usepackage{graphicx}
\usepackage{tabularx}
\usepackage{subfigure}
\usepackage[font=rm]{caption}
\DeclareCaptionType{copyrightbox}
\usepackage{booktabs}
\usepackage{xspace}
\usepackage{enumerate}
\usepackage{enumitem}
\usepackage{shortvrb}
\usepackage{algpseudocode}
\usepackage{mathtools}
\usepackage[mathscr]{eucal}
\usepackage{url}
\usepackage[all]{nowidow}
\usepackage{balance}
\renewcommand{\cite}{\citep}

\newcommand{\metricfont}[1]{{\small\sf{#1}}}
\newcommand{\metric}[1]{\metricfont{#1}}
\newcommand{\MED}{\metric{MED}}

\newcommand{\med}[1]{{\ensuremath\metric{MED}_{\small\metricfont{#1}}}}
\newcommand{\coll}{\mbox{\ensuremath{\mathcal{D}}}}

\newcommand{\argmin}{\operatornamewithlimits{argmin}}

\newcommand{\rhomax}{\rho_{\mbox{\scriptsize{max}}}}

\newcommand{\wand}[0]{\textsc{Wand}{}}
\newcommand{\bmw}[0]{\textsc{Bmw}{}}
\newcommand{\jass}[0]{\textsc{Jass}{}}
\newcommand{\daat}[0]{\textsc{DaaT}{}}
\newcommand{\saat}[0]{\textsc{SaaT}{}}
\newcommand{\taat}[0]{\textsc{TaaT}{}}

\newcommand\gb[1]{$#1$\,GB}

\newcommand{\myurl}[1]{{\url{#1}}}
\newcommand{\mycaption}[1]{\caption{\normalfont{#1}}}

\newcommand{\myparagraph}[1]{~\\\noindent{\textbf{#1.}}~}

\newcommand{\noi}{\noindent}
\newcommand{\mycomment}[1]{}

\newlength{\onedigit}
\newcounter{todocount}
\begin{document}

\title{Efficient and Effective Tail Latency Minimization in 
Multi-Stage Retrieval Systems}


\author{Joel Mackenzie}
\affiliation{
  \institution{RMIT University}
  \city{Melbourne}
  \country{Australia}
}
\email{joel.mackenzie@rmit.edu.au}

\author{J. Shane Culpepper}
\affiliation{
  \institution{RMIT University}
  \city{Melbourne}
  \country{Australia}
}
\email{shane.culpepper@rmit.edu.au}

\author{Roi Blanco}
\affiliation{
  \institution{RMIT University}
  \city{Melbourne}
  \country{Australia}
}
\email{rblanco@udc.es}

\author{Matt Crane}
\affiliation{
  \institution{University of Waterloo}
  \city{Waterloo}
  \country{Canada}
}
\email{matt.crane@uwaterloo.ca}

\author{Charles L. A. Clarke}
\affiliation{
  \institution{University of Waterloo}
  \city{Waterloo}
  \country{Canada}
}
\email{claclarke@gmail.com}

\author{Jimmy Lin}
\affiliation{
  \institution{University of Waterloo}
  \city{Waterloo}
  \country{Canada}
}
\email{jimmylin@uwaterloo.ca}



\setcopyright{none}
\renewcommand\footnotetextcopyrightpermission[1]{} 
\pagestyle{plain}



\clubpenalty=10000
\widowpenalty = 10000


\begin{abstract}
Scalable web search systems typically employ multi-stage retrieval
architectures, where an initial stage generates a set of candidate
documents that are then pruned and re-ranked. Since subsequent stages
typically exploit a multitude of features of varying costs using machine-learned models, reducing
the number of documents that are considered at each stage improves
latency.
In this work, we propose and validate a unified framework that can be
used to predict a wide range of performance-sensitive parameters which 
minimize effectiveness loss, while simultaneously minimizing
query latency, across all stages of a multi-stage search architecture.
Furthermore, our framework can be easily applied in large-scale IR
systems, can be trained without explicitly requiring relevance
judgments, and can target a variety of different efficiency-effectiveness
trade-offs, making it well suited to a wide range of search scenarios.
Our results show that we can reliably predict a number of different
parameters on a per-query basis, while simultaneously detecting and
minimizing the likelihood of tail-latency queries that exceed a pre-specified
performance budget.
As a proof of concept, we use the prediction framework to help
alleviate the problem of tail-latency queries in early stage retrieval.
On the standard ClueWeb09B collection and $31$k queries, we show that our new hybrid system
can reliably achieve a maximum query time of $200$ ms with
a $99.99\%$ response time guarantee without a significant loss
in overall effectiveness.
The solutions presented are practical, and can easily be used
in large-scale distributed search engine deployments with a
small amount of additional overhead.

\end{abstract}




\maketitle


\section{Introduction}
\label{sec-intro}

The competing goals of maximizing both efficiency and effectiveness
in large-scale retrieval systems continue to challenge builders of search systems
as the emphasis in modern architectures
evolves towards multi-stage retrieval~{\cite{p10-query}}.
Many old efficiency problems become new again in the increasingly
complex cascade of document re-ranking algorithms being developed.
For example, research groups can focus on early stage retrieval
efficiency~{\cite{al13sigir,ccl16-adcs,wds16-sigir}}, balancing
feature costs~{\cite{wlm11sigir,cx+14-jmlr}}, or improving the
performance of the learning-to-rank algorithms
~{\cite{msoh13acmtois,alv14-tkde,lno+15-sigir,lno+16-sigir,jyt16-sigir}}.

While great strides have been made in all of these areas, gaps remain
in our understanding of the delicate balance between efficiency and
effectiveness in each ``stage'' of the re-ranking cascade.
One of the most significant limitations preventing further progress
is in training data availability.
While query sets to measure efficiency in various collections are
plentiful, the costs of gathering relevance judgments in order to
measure effectiveness limit the number of topics available for
more detailed trade-off analyses.

In this work we explore how to apply a {\em reference list} framework
~{\citep{wmz10acmtois,tc15,ccm16irj,skc16-tois}} to alleviate this problem.
We leverage the new framework to build machine-learned models 
capable of predicting query response times, candidate set sizes in
early stage retrieval, and algorithm aggressiveness to balance
efficiency and effectiveness on a query-by-query basis.
In particular, we focus on using this unified framework to identify
and reduce {\em tail-latency queries}~\cite{db13-cacm, jk+14-sigir,
sk15-wsdm, hk+16-tweb}, i.e., those with unusually large response time.
We explore three important research questions:

\smallskip

\noi{\bf Research Question 1 (RQ1):} {\emph{
What is the best way to use reference lists to accurately
perform dynamic per query parameter predictions in early stage
retrieval?
}} 

\smallskip

\noi{\bf Research Question 2 (RQ2):} {\emph{
What is the relationship between tail-latencies and index traversal
algorithm, and can our new prediction framework be used reliably provide
worst case guarantees on first-stage query efficiency?
}} 

\smallskip

\noi{\bf Research Question 3 (RQ3):} {\emph{
What combination of predictions will lead to efficient first-stage
retrieval, minimizing the number of candidate documents exiting
the first stage (and thus making later stages more efficient),
and also minimize effectiveness loss in final stage re-ranking?
}} 

\smallskip
\noindent In answering these questions, our research contributions include:

\begin{enumerate}[leftmargin=*]
\item A unified framework that can be used to predict a wide variety
of performance-sensitive parameters in multi-stage retrieval systems.
\item A pragmatic, yet highly effective solution to tail-latency query 
minimization that can easily be implemented in large-scale retrieval
systems, and provide worst case performance guarantees on performance.
\item A pathway to more fine-tuned per-query optimization techniques,
and the tools necessary to implement and test systems leveraging these
ideas.
\end{enumerate}

\noindent We achieve these goals using three ideas.
First, we exploit the idea of query difficulty prediction~{\citep{cyt10-qpb}}
and static pre-retrieval features to build a unified prediction framework.
Next, we explore the relationship between the number of documents
returned in a top-$k$ candidate set and the index traversal algorithm.
Three different index traversal algorithms have been commonly used:
document-at-a-time (\daat{}), term-at-a-time (\taat{}), and
score-at-a-time (\saat{}).
A recent paper by {\citet{cc+17-wsdm}} performed a comprehensive comparison
of state-of-the-art \daat{} and \saat{} algorithms and found that both
approaches have advantages and disadvantages.
In this work we look at a simple index mirroring approach which
selectively uses the best algorithm based on a series of
pre-retrieval predictions.
Finally, the efficiency predictors are integrated with an effectiveness
loss minimization prediction.
Together, this series of ``Stage-0'' pre-retrieval predictions
produces a pipeline that maximizes efficiency and effectiveness in a
multi-stage retrieval system, and is capable of achieving 
$99.99\%$ response time guarantees when using a worst case
running time of $200$ ms on a commonly used web collection.


\section{Background and Related Work}
\label{sec-background}
\myparagraph{Efficient Query Processing}
Efficient query processing can be attained through a range of index
organizations and traversal strategies based on the {\it inverted
index} data structure~\cite{zm06compsurv}.
Document-at-a-time (\daat{}) query processing relies on postings
lists being sorted in ascending order of the document identifiers.
At query time, a pointer is set at the beginning of each postings
list.
Once the current document has been evaluated, the pointers are
forwarded to the next document in the lists.
An efficient method for {\it disjunctive} \daat{} processing is the
{\it Weak-AND} (\wand{}) algorithm~\cite{zchsz03cikm}.
In order to support \wand{} traversal, the upper-bound score that
term $t$ can contribute to any given document must be pre-computed
and stored in the index ($U_t$).
At query time, \wand{} uses the lowest-scoring heap document as a
{\it threshold}.
When selecting the next document in which to score, \wand{} will only
select a document in which the sum of the $U_t$ scores is larger than
the heap threshold.
The advantage of \wand{} is that documents that are not able to make
the final top-$k$ results are able to be safely ignored, making it
highly efficient.
Although originally aimed for traversing on-disk indexes, \wand{} has
been proven to be efficient in-memory on many
occasions~\cite{fj+11-pvldb, pcm13-adcs, mcc15-adcs, cc+17-wsdm,
nt11-sigir, al13sigir}.

\citet{dt11-sigir} (and at a similar time,~\citet{ccv11-icde})
explored an improved version of \wand{} named {\it Block-Max WAND}
(\bmw{}).
The key observation in \bmw{} is that since many index compression
algorithms are block-based~\cite{lb-spe15, jz08-www}, skipping can be
achieved at the block level, thus saving an entire block
decompression.
In order the facilitate this skipping, the $U_t$ score is computed
for every {\it block} in each postings list, known as the $U_{b,t}$
score.
When a pivot document is found (by summing the $U_t$ scores until the
threshold is exceeded), the local block score is then used to refine
the estimated score, that is, the sum of the $U_{b,t}$ scores is
computed.
If this sum still exceeds the threshold, then the pivot document is
scored. Otherwise, a new pivot is selected.
Additional gains from \bmw{} are achieved through an improved
skipping function that identifies if the current block configuration
could not contain a document with a score above the threshold.
If this condition is met, a new pivot is selected that {\it may}
contain enough weight to enter the top-$k$ heap.
Further enhancements to \bmw{} have been made in the literature,
usually by using additional auxiliary structures that provide a
quicker search time while using additional space, or using hybrid
indexes~\cite{dns13-wsdm, cr13-sigir,pmc14-sigir}.

Another entirely different method for top-$k$ query processing is the
term-at-a-time (\taat{}) and the closely related score-at-a-time
(\saat{}) approach.
Term-at-a-time processing opts to process an entire postings list
before moving onto the next list.
Clearly, an additional data structure must be kept to store the
partially accumulated scores while processing the lists.
{\citet{akm01-sigir}} made the observation that the term weight for
any given document $w_{d,t}$ could be pre-computed and stored, rather
than the term frequencies ($f_{d,t}$).
Since the $w_{d,t}$ are typically floating point numbers, they are quantized
into integer values to facilitate compression~\cite{akm01-sigir}, the range of
which impacts both effectiveness and efficiency~\cite{mt13-cikm}.  For
score-at-a-time processing, each postings list is sorted by {\it
decreasing} impact score, which allows the most high scoring
documents for each term to be processed first, and can allow for
early-termination without sacrificing effectiveness.
Recently,~\citet{lt15-ictir} introduced \jass{}, a modern \saat{}
algorithm which can be used for {\it anytime} retrieval, making it
suitable for use in time-constrained environments and for controlling
tail latencies.

Finally, some optimizations can be generalized to all index
structures.
For example, many compression algorithms have been proposed in the
literature~\cite{lb-spe15,t14-adcs,tl16-adcs,jz08-www} which are
often applicable to frequencies, (quantized) document weights, and
DocIDs.
Another general improvement is to apply a special ordering to the
DocID space~\cite{sd10-www, nt11-sigir,ak14-lsdsir}.
Assignment strategies such as lexicographically sorting the DocIDs by
the corresponding {\it URL} has been shown to improve both the
compression rate, and reduce the query latency~\cite{nt11-sigir,sil07}.

\myparagraph{Tail Latencies}
A tail-latency query is an ``outlier'' query whose response time occurs above
the $n$th percentile, where $n$ is a large value such as $95$, $99$,
or even $99.99$~\cite{yh+15-sigir, sk15-wsdm}.
As collections grow larger, systems must scale accordingly. As systems become
more complex, the probability
of tail latencies occurring also increases~\cite{db13-cacm}, particularly for
distributed architecture where end-to-end latency is often bound by the slowest component.
Tail latencies can be addressed through either hardware or software optimizations, or both.
For example, replicating and partitioning
collections~\cite{db13-cacm, kccm16-sigir, gf14-wsdm, kccm16-irj}
allows effective load balancing which can minimize tail-latency queries.

Previous work has attempted to reduce tail latencies in a range of
different contexts.
{\citet{jk+14-sigir}} focus on $99$th percentile tail-latency queries at the
level of a single {\it Index Server Node} (ISN) by predicting long
running queries, and running them in parallel.
Queries that are {\it not} predicted as long running are simply ran
sequentially, which avoids the overhead cost of parallalization.
Another recent work targets extreme tail latencies (ie, at the
$99.99$th percentile)~\cite{sk15-wsdm,hk+16-tweb}.
This target is achieved through {\it Dynamic, Delayed, Selective}
(DDS) prediction.
DDS prediction works as follows.
First, a new query is ran for a short time, such as $20$ms, and
dynamic features are collected from this initial processing.
Then, new dynamic features (and, some additional static features)
are used to predict whether the query is a long running query.
If so, or if there is reasonable uncertainty (based on the predicted
error), then the query will be accelerated using parallelization.
The prediction error is then used to improve coverage of midpredicted
true long running queries.

Beyond the tail latency in ISNs, DDS also reduces the latency of the
aggregator node, which aggregates the results from the multiple ISNs
before reporting them to the user.
~\citet{yh+15-sigir} also address the problem of aggregating
information from ISNs, but this is orthogonal to our work, which
focuses on the processing at an ISN, and not at the aggregation node.

\begin{figure}[t]
\centering
\includegraphics[width=\columnwidth]{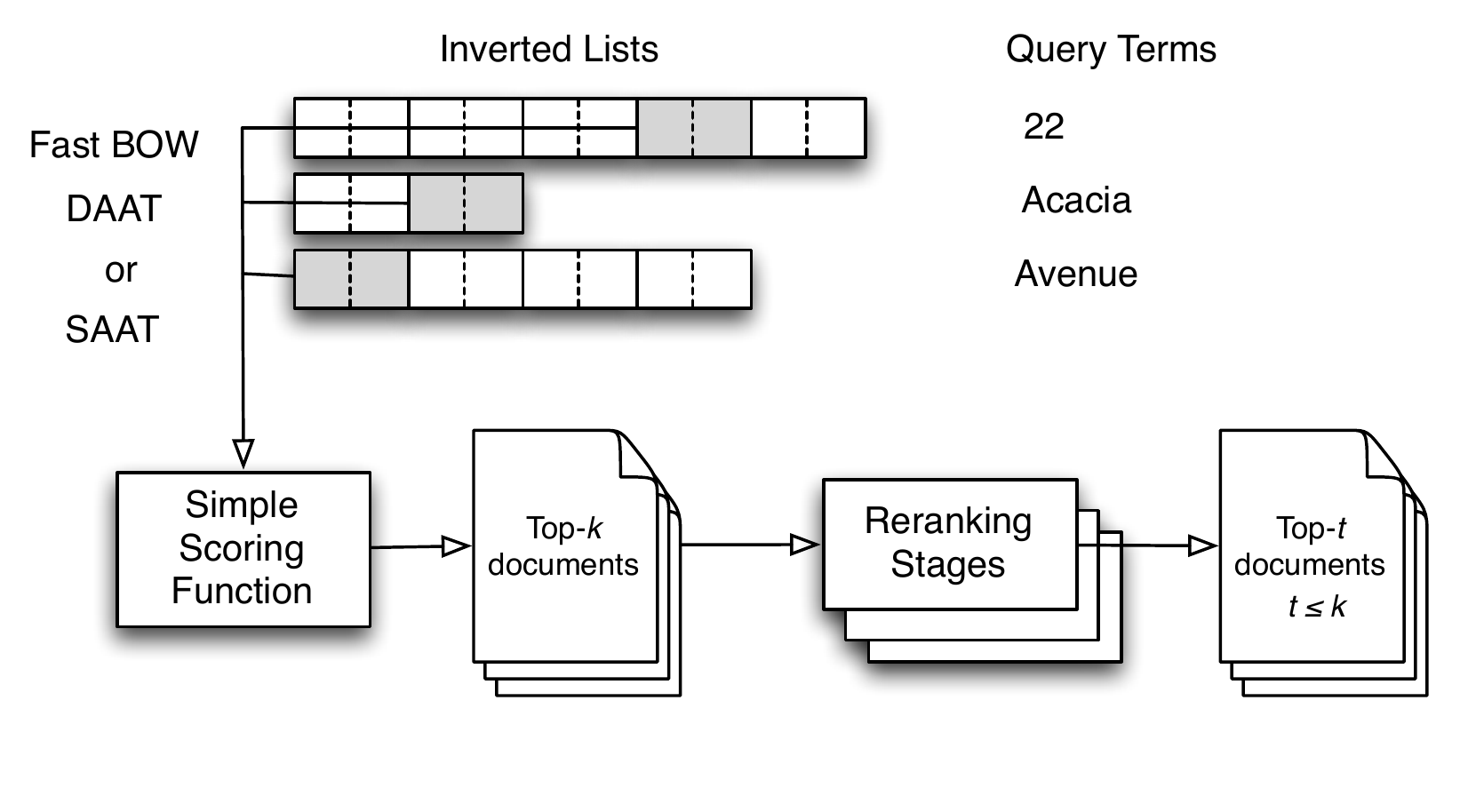}
\mycaption{Architecture of a typical multi-stage retrieval system.
Queries are first processed using an efficient bag-of-words
processing algorithm. The initial candidate set of $k$ documents
then undergoes a series of re-ranking stages where the candidate
pool is shrunk, and more expensive learning-to-rank algorithms
are used to produce a final set of top-$t$ documents to return
to the user, where often $t \ll k$.
}
\label{fig-mstage}
\end{figure}

\myparagraph{Multi-stage Search Architectures}
Multi-stage retrieval has become the dominant model in modern web search
systems~{\citep{p10-query,czc10-wsdm,mso13irj,msoh13acmtois,al13irj,al13sigir}}.
In this approach, a set of candidate documents are generated that are
likely to be relevant to a query, and then in one or more stages,
the document sample is iteratively reduced and reordered
using a series of increasingly expensive machine learning
techniques.
Since re-ordering can be computationally expensive and is sensitive
to the number of documents that must be reordered, minimizing the
size of the candidate set is an important problem
~{\citep{czc10-wsdm,mso13irj,tmo13-wsdm}}.

Figure~{\ref{fig-mstage}} exemplifies a typical multi-stage retrieval
architecture.
A fast bag-of-words processing algorithm produces a top-$k$ candidate
set.
This initial set of documents is then re-ranked one or more times
using a learning-to-rank algorithm to produce a final output set
of $t$ documents, where $t \le k$, and can be $t \ll k$ in some
configurations.

Efficiency matters at all stages of the process.
{\citet{kdf13-kdd}} showed that every $100$ ms boost in overall
search speed increases revenue by $0.6\%$ at Bing.
So, even small gains in overall performance can translate to tangible
benefits in commercial search engines.
Efficiency remain an important problem in multi-stage retrieval with
papers focused on cascaded ranking~{\citep{p10-query,cx+14-jmlr}},
and early exit optimizations~{\citep{czc10-wsdm,dbc13-ecir}.
Recently,~\citet{wds16-sigir} proposed a fast candidate generation framework
which opts to build a two-layer index. The bottom layer is the standard inverted
index, and the top layer is a single or dual-term auxillary structure which
stores a subset of the bottom layer documents, sorted by impact score. At
query time, a prefix of the top layer is accessed, which generates a set of
candidate documents. Then, the most promising of these candidate documents
has its partial scores updated by accessing the lower layer of the index (to
achieve a more accurate score). Finally, the top-$c$ candidates are selected
and passed onto the next stage of the cascade. We do not consider this
generation framework as it provides approximate results, but note that it 
can be directly applied to our existing \bmw{} ISN to improve efficiency (with
some small loss in effectiveness). We leave this as future work.


\myparagraph{Effectiveness Evaluation in Multi-Stage Retrieval}
One obvious question arises when trying to measure trade-offs in
multi-stage retrieval systems.
The simplest approach is to simply make changes to the system,
and re-compute a standard information retrieval metric such as
average precision (AP), expected reciprocal rank (ERR),
normalized discounted cummulative gain (NDCG),
or rank biased precision (RBP) on the last stage result~\cite{cmzg09cikm, mz08acmtois}.
However, this is unwieldy in practice, as it can be very difficult
to identify exactly what changes are resulting in effectiveness
differences.

A better approach is to compute intermediate results at different
stages of re-ranking, and measure the differences between the two.
For example, in a simple two-stage system, we could generate the
top-$k$ list for both stages and somehow measure the similarity
or difference between the two runs.
We refer to this as a {\em reference list} comparison.
For example, we could just compute the {\em overlap} between
the two lists, and this methodology is still commonly used
in recent work~{\citep{wds16-sigir}}.
But in practice, this approach does not properly capture importance
of rank position in the two lists.
To alleviate this problem, {\citet{wmz10acmtois}} proposed rank-biased
overlap (RBO).
This is a non-conjoint list comparison metric that places more importance
on the loss of higher ranking items in a list than lower ranking ones.

The goals of RBO were taken one step further by {\citet{tc15}} in
the metric {\em Maximized Effectiveness Difference} ($\MED$) where
the exact gain function used to compute the difference can depend
on any utility-based evaluation metric, such as ERR, DCG, or RBP.
Furthermore, $\MED$ has the additional advantage that if partial
judgments are available for any of the queries, the information
can be used directly for the final comparison. 
%
%
\noindent Informally, $\MED$ answers the following question: 
given an effectiveness metric and two ranked lists,
$\coll^a$ and $\coll^b$, what is the maximum difference in the
effectiveness scores between the two lists?
\citet{tc15} define variants of MED for many standard
retrieval metrics, including
average precision (MED-AP), expected reciprocal rank (MED-ERR),
normalized discounted cumulative gain (MED-NDCG),
and rank biased precision (MED-RBP). We refer the reader to the work of
~\citeauthor{tc15} for
the formal definition of $\MED$.
In this paper we employ MED-RBP with a decay value of $0.95$ 
(MED-RBP$_{0.95}$) as our primary difference measure.

Other approaches to defining reference lists have been studied
recently by {\citet{skc16-tois}}.
Their approach is orthogonal to the one taken in this work.
The relationship between how best to construct ground truth runs and
measure the similarity between two non-conjoint lists remains a
fruitful area of future research in the IR community, but is beyond
the scope of this work.

\section{Methodology}
\label{sec-methods}

In order to build our prediction framework, we need to account for
several issues.
First, we need a ground truth which represents an idealized last
stage run over a large corpus of queries.
This idealized last stage represents the reference list for which
all comparisons can be made.
In order to build a plausible reference list, we adopt the methodology
of {\citet{ccm16irj}}.
The 2009 Million Query Track (MQ2009) query set was used to perform
both training and testing.
We filtered this query set by removing single term queries (which can
be answered extremely efficiently by taking the first $k$ documents
from the relevant postings list of the impact-ordered ISN).
Following~\citeauthor{ccm16irj}, we use the {\tt uogTRMQdph40} run as a
reference list, as it was one of the highest performing runs across
the evaluated query set, and had results for all of the queries in
the collection.
In addition, we filtered out $905$ queries which reported a
MED-RBP$_{0.95}$ score greater than $0.5$ when applying the fixed-$k$
early stage (with $k = 10{,}000$), as
these results show a clear mismatch between the early and late
stages we are presenting. 
After filtering, we retain a set of $31{,}642$ MQ2009 queries.
The first $50$ queries are held out for final effectiveness validation
since these queries correspond to the queries in the 2009 TREC
Web Track, and a full set of relevance judgments are available.
For all predictions, queries were randomly assigned to $10$ folds,
and standard $10$ fold cross validation was performed to produce
the query predictions.

We use only MED-RBP$_{0.95}$ with a small target
threshold of $\epsilon=0.001$ for all experiments as we wish to
aggressively minimize effectiveness loss.
\citeauthor{ccm16irj} showed that other common utility-based
metrics could also easily be used such as ERR and DCG, and achieve
similar results in their experiments, but we do not explore that
option in this work.

\myparagraph{Experimental Setup}
All experiments were executed on an idle 24-core Intel Xeon E5-2690
with {\gb{512}} of RAM hosting RedHat RHEL v7.2.
ATIRE~\cite{tjc12-osir} was used to parse and index the ClueWeb09B
collection, which was stopped using the default Indri stoplist, and
stemmed using an {\it s}-stemmer.
Timings were conducted on an appropriate
\bmw{}\footnote{\url{http://github.com/JMMackenzie/Quant-BM-WAND}} or
\jass{}\footnote{\url{http://github.com/lintool/JASS/}} index, which
use QMX compression~\cite{t14-adcs,tl16-adcs} and the BM25 scoring model.
Each query is processed $5$ times, and the average of the $5$ runs is
reported.

\myparagraph{Prediction Framework}
Recently, {\citet{ccl16-adcs}} described an effective approach
to dynamically predicting $k$ while minimizing the effectiveness
loss.
Their key idea was to use the
reference list methodology described above to build ground truth
labels to train a classifier.
However, their approach has a few drawbacks.
First, the cascade classifier they described is interesting but
unconventional in that it requires multiple predictions to be made,
depending on the final $k$.
Fewer predictions are required for small $k$, but up to $8$
independent predictions are required for large $k$.
Secondly, the problem they describe is really a regression problem in
practice.
Using regression allows an exact $k$ to be predicted instead of an
approximate cutoff, which translates into fewer documents being
re-ranked in later stages of the retrieval system.

Commonly, regression methods estimate the conditional expectation of
a target dependent variable $y$ given the independent variables (or
features) $\mathbf{x}$.
This implies that the method approximates the average value of the
dependent variable when the independent variables are fixed.
Given training data of the form $(\mathbf{x_1},y_1), \dots,
(\mathbf{x_n}, y_n)$ methods based on least squares try to optimize
the loss function $L(\mathbf{x}, y)=\frac{1}{n}\sum_{i=1}^n
\frac{1}{2} (\mathbf{x_i} - y_i)^2$, which results in a good
estimator for the mean $E[y|\mathbf{x}]$.

\begin{figure}[t]
\centering
\includegraphics[width=\columnwidth]{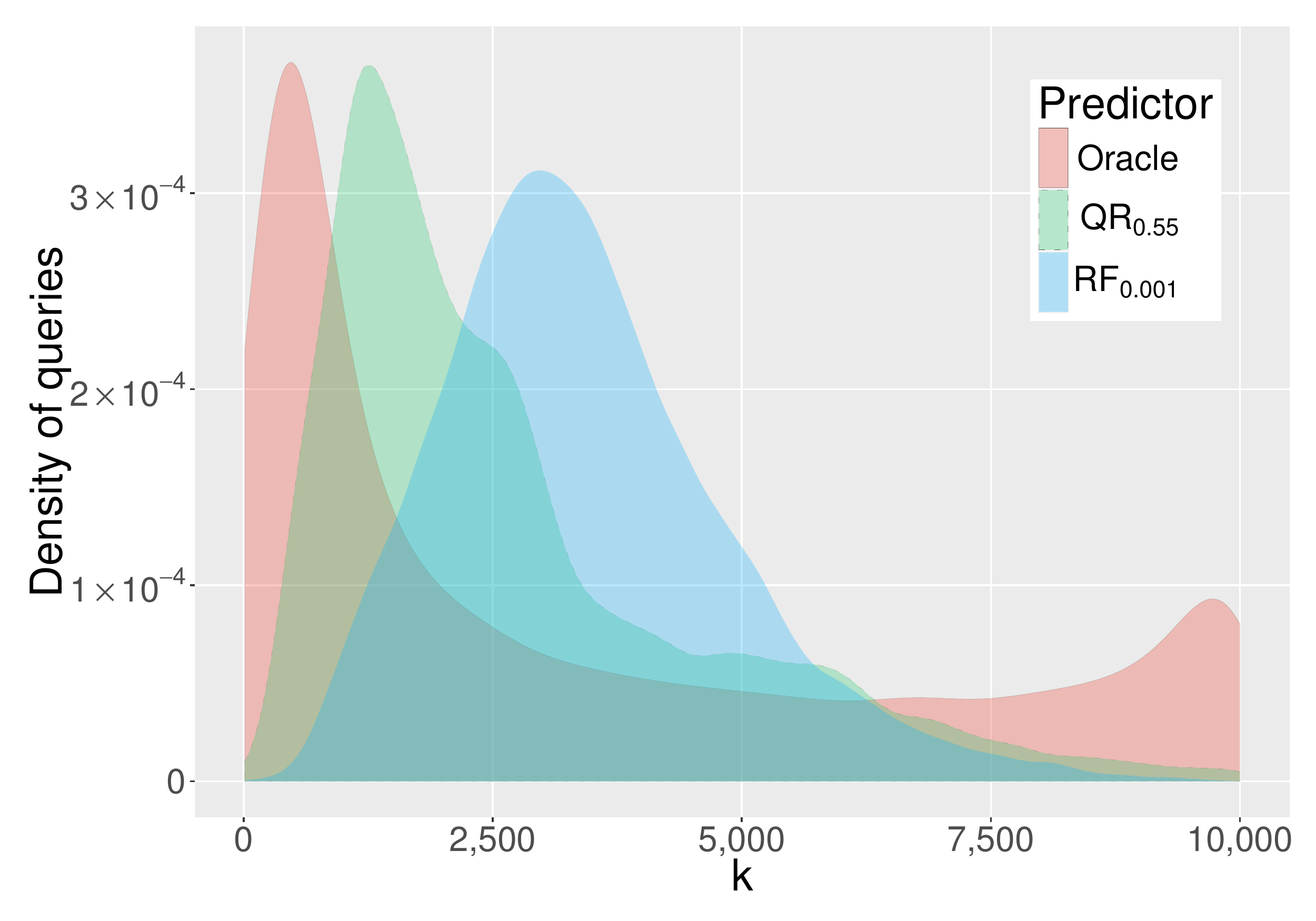}
\mycaption{A comparison of the distributions of actual $k$ versus 
predicted $k$ when using a Random Forest regression and a
Quantile Regression in first stage
retrieval for the $31{,}642$ queries from the
MQ2009 TREC Task. Note that the Random Forest uses a
training value of $\epsilon = 0.001$, whereas the best-fit distribution
for the Quantile Regression was $\tau = 0.55$ for $k$.
}\label{fig-kdist}
\end{figure}

So, the obvious way to reproduce their work is to use a similar
feature set, and compute the exact $k$ needed for each query
that achieves a very small expected $\MED$ loss, say, $\epsilon < 0.001$,
and use a random forest to produce the predictions.
When we build this training set, one immediate problem becomes
apparent -- the ground truth labels do not follow a standard
distribution, but an out-of-the-box regression algorithm {\em does}.
Figure~{\ref{fig-kdist}} shows three different distributions -- the
true distribution of $k$ in the ground truth set (Oracle), the
random forest prediction ($\mbox{RF}_{0.001}$), and a quantile
regression prediction ($\mbox{QR}_{\tau}$), which is described
now.

A pitfall of standard regression methods is that they may become
unstable under heavy-tailed distributions due to the dominant effects
of outliers, or more precisely, when samples from the tail of the
distribution have a strong influence on the mean.
How to cope with this problem has been studied in the context of
\emph{robust estimation}.
These estimators embody a family of methods designed to be more
resilient to the data generation process by not following the underlying
assumptions behind the regressor; in the context of least squares,
this would be errors being uncorrelated and having the same variance.

One simple way of dealing with the outlier problem is \emph{quantile
regression} which estimates either the conditional median or other
quantiles of the response variable.
If $y$ has a cumulative distribution of $F_y(z)=p(y \leq z)$ then the
$\tau$-th quantile of $y$ is given by $Q_y(\tau)=F_y^{-1} = \inf\{z :
F_y(z) \geq \tau\}$.
To learn a regressor that minimizes a $\tau$ value, we define the
loss function $\xi_\tau(y) = y(\tau - \mathcal{I}\{y < 0\})$ where
$\mathcal{I}\{\cdot\}$ is the indicator function.
Therefore, $\tau$-th quantile regression estimates the conditional
$\tau$-th quantile $F^{-1}_{y|\mathbf{x}} (\tau)$, or we want an
estimate $\hat{f}_\tau$ such that $p(y < \hat{f}_\tau(\mathbf{x})) =
\tau$:
\begin{eqnarray}
\hat{f}_\tau = \argmin_{f \in \mathcal{F}_\tau} \sum_{i=1}^n \xi_\tau(y_i - f(\mathbf{x_i})) = \\
 \argmin_{f \in \mathcal{F}_\tau} \left[ (1-\tau) \sum_{y_i < f(\mathbf{x_i})} \lvert y_i - f(\mathbf{x_i})\rvert + \tau \sum_{y_i \geq f(\mathbf{x_i})} \lvert y_i - f(\mathbf{x_i}) \rvert\right] \;,
\end{eqnarray}
where $ \mathcal{F}_\tau$ is a predetermined class of functions.

A robust regression method is \emph{random forests} (RF) which build
several decision trees using attribute bagging.
In a nutshell, the algorithm samples with replacement the training
data $B$ times and trains several decision trees $f_b$ using only each
portion of the data.
The final prediction for an incoming new query is averaged from all
the regressors $\hat{f}=\frac{1}{B}\sum_{i=1}^B f_B(\mathbf{x})$.
Subsampling has the practical effect of decreasing the variance of
the model, without increasing its complexity, given that even if the
predictions of a single tree are highly sensitive to noise, the
average of many trees is not, as long as the trees are not
correlated.
Bootstrapping achieves this effect by training each tree with a
different randomized subsample.

When the individual trees $f_b$ are learned, the building procedure
has to create tree nodes that branch the data down the tree; in order
to reduce the model variance, only a few features are candidates for
splitting at each round.
This mitigates the effect that happens when, if just a few features
are very strong predictors for $y$, these features will be selected
in many of the $B$ trees, which will become correlated.
Given their resilience to noise and outliers, random forest were the
best out-of-the-box regressors for the task of predicting cut-off
values and query response times, surpassing in effectiveness many
other candidates such as kernel ridge regression, Gaussian
(regression) processes among others.

We deploy the quantile regression within the same tree framework
using gradient boosting regression trees (GBRT).
In this case, each tree re-fits the training data using the residuals
(gradients) of the training data with respect to the $\xi_{\tau}$
loss function, and a pre-tree weight is calculated using line search.
The final decision is a linear combination of the weighted prediction
of the tree ensemble.

We used a similar set of features as~\citet{ccl16-adcs}. These features
are based on a aggregating statistics for each postings list 
(such as maximum scores, harmonic/arithmetic mean/median
scores, and so on) from a range of similarity functions, along with query
specific features such as query length, max score of query terms, and many more.
In addition to the TF$\cdot$IDF, BM25 and query likelihood used 
in~\cite{ccl16-adcs}, we also build features using Bose-Einstein, DPH, and DFR
similarity functions~\cite{ar02-tois}. We also added the geometric mean as an
aggregation statistic for each of these similarity functions.
We use a total of $147$ features in this work.

\myparagraph{Tail-Latency Queries in {\daat{}} Traversal Algorithms}
This improved approach to predicting $k$ in first stage retrieval
is a promising first step to achieving for efficient results
without sacrificing effectiveness.
However, assuming that the performance of \wand{}-based algorithms
in the first stage is a function of $k$ may not be correct in practice,
and other recent work~{\citep{cc+17-wsdm}} provides
persuasive evidence that this assumption is not true in practice.
{\citeauthor{cc+17-wsdm}} showed that when using \wand{} and
\bmw{}, tail-latency queries can occur at any $k$ cutoff,
making performance guarantees hard to enforce in production
systems. 

The alternative to using \wand{} or \bmw{} in the first stage
retrieval is to use a \saat{} algorithm such as \jass{}.
Unfortunately, this is not an entirely satisfactory answer either
as most of the performance gains in \jass{} come from using
aggressive early termination, which can hurt effectiveness when
the number of documents that must be passed into the next stage
must also be minimized.
So rank safety is yet another confounding factor.
\daat{} and \saat{} processing algorithms can sacrifice effectiveness for
efficiency by relaxing the rank-safety constraint.
For example, \jass{} allows a parameter $\rho$ to be set which bounds
the maximum number of postings to score per query,
and variants of \wand{} can use a parameter $\theta$
(or sometimes $F$) which induces more aggressive skipping
during postings list traversal.
So, there is a trade-off between retrieval depth $k$ and
rank safety in a pure efficiency sense.
This relationship was previously explored by {\citet{tmo13-wsdm}},
who also used a query difficulty prediction framework to
solve the problem.
We build on this idea in this work, but also account for the
fact that using only \wand{} based algorithms can still result
in poorly performing tail-latency queries.
We can see that boosting $\theta$ alone does indeed make \bmw{} faster
in Figure~{\ref{fig-tail}}, but the tail-latency queries remain.

\begin{figure}[t]
\centering
\includegraphics[width=\columnwidth]{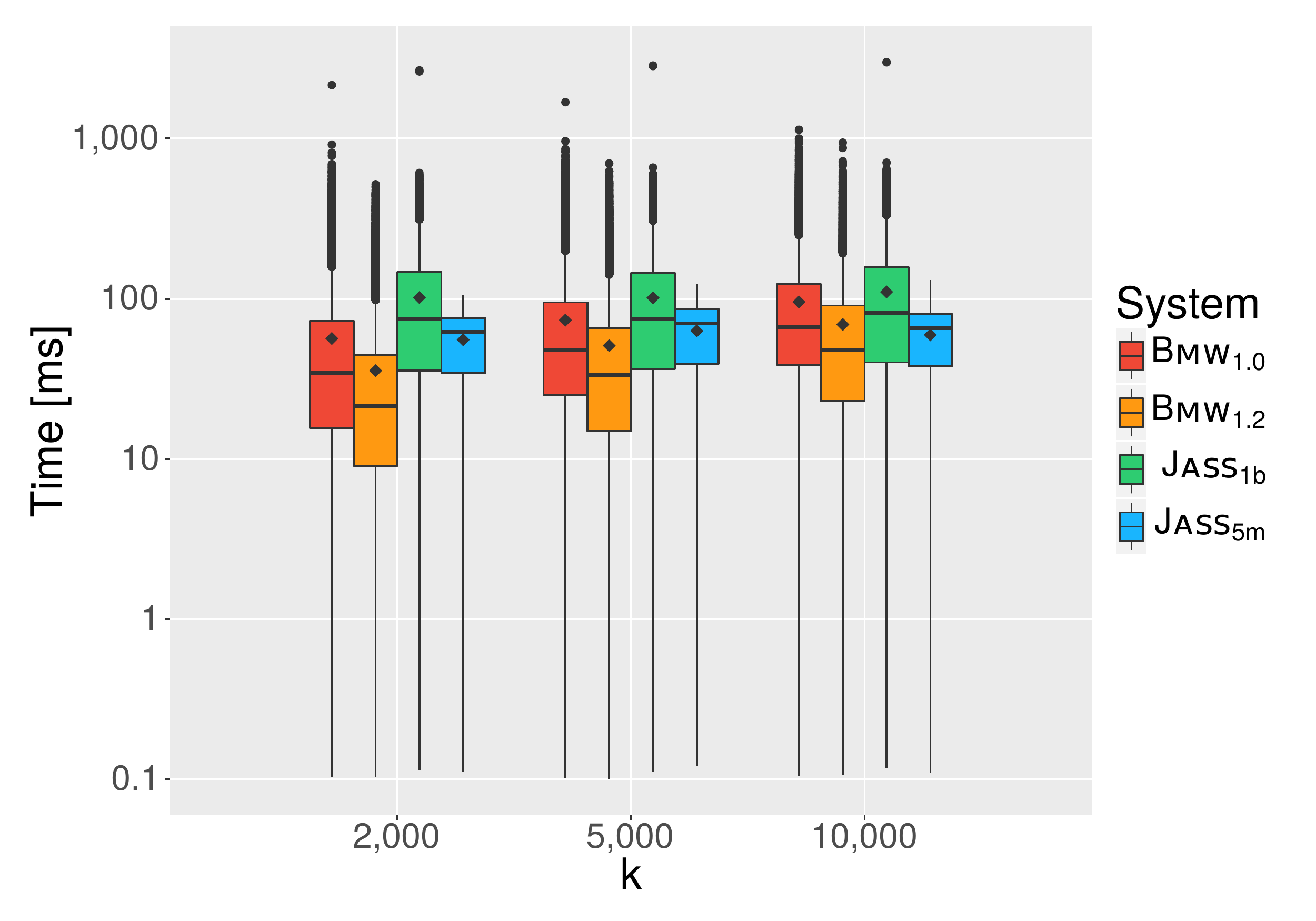}
\mycaption{Efficiency comparison of the $31{,}642$ queries from the 
	MQ2009 TREC Task using both aggressive and exact versions of \bmw{} and \jass{}.
Subscripts denote the aggression parameters ($\theta$ for \bmw{} and $\rho$
for \jass{}).
}\label{fig-tail}
\end{figure}

So, our next task is to explore the likelihood of tail-latency queries when
using the MQ2009 topic set.
{\citet{cc+17-wsdm}} recently did a comparative analysis using the
UQV~\cite{bmst16-sigir} query set and the ClueWeb12B document
collection with fixed values of $k$.
We reproduce their work here across our own query set and fixed $k$
values.
Figure~{\ref{fig-tail}} shows the breakdown of all $31{,}642$ queries
across a number of fixed values of $k$, selected as appropriate sizes
for an LtR system~\cite{mso13irj}.
Similar to~\citeauthor{cc+17-wsdm}, we observe that the exhaustive
\bmw{} algorithm is superior to the exhaustive \jass{} algorithm, but
the heuristic \jass{} traversal (with the recommended $10\%$ heuristic)
eliminates all tail-latency queries.
On the other hand, the aggressive \bmw{} traversal does improve the
mean and median times, but does not reduce the likelihood of tail-latency
queries.
Note that we selected the value for the heuristic, $\theta = 1.2$,
based on other work that shows that more aggressive approaches
result in reduced effectiveness~\cite{mto12-sigir, ccm16irj}.
It is also noteworthy that the exhaustive \bmw{} traversal has a
faster median time than the aggressive \jass{} traversal when $k \leq
5{,}000$.

\begin{table}[t]
\begin{center}
\resizebox{0.48\textwidth}{!}{
\begin{tabular}{cccccc}
\toprule
& \bmw{}$_{1.1}$ & \bmw{}$_{1.2}$ & \bmw{}$_{1.3}$ & \jass{}$_{1b}$ & \jass{}$_{5m}$\\
\midrule
\bmw{}$_{1.0}$  & $86.0$ & $61.7$ & $46.6$ & $56.2$ & $16.7$\\
\bmw{}$_{1.1}$  & - & $67.4$ & $49.7$ & $53.1$ & $18.3$\\
\bmw{}$_{1.2}$  & - & - & $68.6$ & $42.3$ & $24.2$ \\
\bmw{}$_{1.3}$  & - & - & - & $31.4$ & $26.7$\\
\jass{}$_{1b}$ & - & - & - & - & $8.0$\\
\bottomrule
\end{tabular}
}
\end{center}
\mycaption{The percentage overlap of queries that fall in 95th percentile
	efficiency band for $k = 2{,}000$. 
Clearly, making \bmw{} more aggressive may improve timings, but the tail
queries are generally similar regardless of the aggression parameter $\theta$.
On the other hand, it is less common for \jass{} and \bmw{} to have overlapping
tail queries, especially when a non-exhaustive $\rho$ value is used.}
\label{tbl-overlap}
\end{table}

Additionally, we do a simple overlap analysis on the $95$th
percentile tail-latency for each algorithm to determine whether each system
has similar tail-latency queries.
Table~\ref{tbl-overlap} shows the percentage of the tail-latency queries that
overlap between each system, where $k = 2{,}000$.
Exact \jass{}, exact \bmw{} and aggressive \bmw{} tend to share similar
tail-latency queries.
However, we note that the aggressive \jass{} traversal tends to share
only a small percentage of the tail-latency queries that occur in the other
systems.
This provides further motivation for our proposed hybrid ISN index
configuration.

In light of this new evidence, a pragmatic hypothesis emerges:
Can we somehow combine the best properties of \jass{} and \bmw{}
to create a hybrid approach that captures the best of both
worlds?

\section{Approach}
\label{sec-approach}

\myparagraph{Problem Definition}
First, we define the problem. 
Given a query $q$, a series of re-ranking stages $R$, and a target
evaluation metric $M$ for the final stage, how can we select both $k$
and the processing algorithm $A$ for the initial (bag-of-words) stage
such that $k$, processing time $t$, and effectiveness loss $L$ are
minimized?

\myparagraph{Untangling the objectives}
Our first goal is to untangle the objectives, and describe a unified
methodology to satisfy all of the constraints in a principled way.
We draw inspiration from all of these recent studies.
We still want to minimize $k$ as the performance of later stage
re-ranking algorithms is sensitive to the number of documents, and
we also want to provide performance guarantees on the running time of
the first stage ranker.
The key observation that pulls these seemingly different objectives
together is that the classification approach described by
{\citeauthor{ccl16-adcs}} actually used classic {\em query hardness}
features for the learning model~{\citep{cyt10-qpb,mto12-sigir,tmo13-wsdm}}.
Furthermore, the {\MED} approach allows many more queries to be
used for training than methods which require a full set of relevance
judgments to be available in order to minimize effectiveness loss.
So, we explore the possibility of using a single predictive framework
to minimize all three constraints in a unified way.

\myparagraph{System Architecture} 
The first major difference in our approach is that we opt to build a
hybrid architecture.
Work on distributed IR has shown that an effective approach to
scaling is to replicate popular ISNs~\cite{db13-cacm, kccm16-sigir,
gf14-wsdm, kccm16-irj}.
Here, we assume that we can build ISNs that are optimized for
different types of queries.
In other words, when we build replicas, we may opt to build a
document-ordered index (appropriate for \daat{} traversal), or an
impact-ordered index (appropriate for \saat{} traversal).
This idea is key to our novel framework: Selecting algorithm $a \in
A$ actually refers to selecting an ISN to process the query which is
configured to run algorithm $a$, and ISN selection is already a
common problem in distributed search
architectures~{\citep{db13-cikm,sk15-wsdm}}.
In practice, our ``Stage-0'' predictions would be performed by the
resource selection process in a large-scale distributed IR system. 

\begin{algorithm}[t]
    \SetKwInOut{Input}{Input}
    \SetKwInOut{Output}{Output}

\caption{Candidate generation pipeline based on predicting $k$}
\Input{A query $q$, 
a regressor $\mathcal{R}_k$ that predicts the required $k$ for $q$,
a regressor $\mathcal{R}_\rho$ that predicts the required $\rho$ for \jass{}
up to a maximum $\rho$ value $\rhomax$, and a $k$-threshold $T_k$}
\Output{A set of candidate documents, $C$}
$C \leftarrow \varnothing$\\
$P_k \leftarrow \mathcal{R}_k(q)$\\
\eIf{$P_k > T_k$}
{
  $P_\rho \leftarrow \mathcal{R}_\rho(q)$\\
  $C \leftarrow $ ISN$_{\jass{}}$$(q, P_k, P_\rho)$\\
}
{
  $C \leftarrow $ ISN$_{\bmw{}}$$(q, P_k)$\\
}
return $C$
\label{alg-1}
\end{algorithm}

\begin{algorithm}[t]
    \SetKwInOut{Input}{Input}
    \SetKwInOut{Output}{Output}

\caption{Candidate generation pipeline based on predicting both $k$ and run time}
\Input{A query $q$, 
a regressor $\mathcal{R}_k$ that predicts the required $k$ for $q$,
a regressor $\mathcal{R}_\rho$ that predicts the required $\rho$ for \jass{}
up to a maximum $\rho$ value $\rhomax$,
a $k$-threshold $T_k$, 
a regressor $\mathcal{R}_t$ that predicts the running time of $q$, 
and run-time threshold $T_t$}
\Output{A set of candidate documents, $C$}
$C \leftarrow \varnothing$\\
$P_k \leftarrow \mathcal{R}_k(q)$\\
\eIf{$k > T_k$}
{
  $P_\rho \leftarrow \mathcal{R}_\rho(q)$\\
  $C \leftarrow $ ISN$_{\jass{}}$$(q, P_k, P_\rho)$\\
}
{
  $P_t \leftarrow \mathcal{R}_t(q)$\\
  \eIf{$P_t > T_t$}
  {
    $P_\rho \leftarrow \mathcal{R}_\rho(q)$\\
    $C \leftarrow $ ISN$_{\jass{}}$$(q, P_k, P_\rho)$\\
  }
  {
    $C \leftarrow $ ISN$_{\bmw{}}$$(q, P_k)$\\

  }
}
return $C$
\label{alg-2}
\end{algorithm}

\myparagraph{Hybrid Approaches}
Based on several observations about the relative performance of
\jass{} and \bmw{}, we are now in a position to describe a few
different hybrid approaches to query processing.
Our goal is to limit the disadvantages of each traversal algorithm,
and exploit the desirable properties.
Several different variations were used in our preliminary experiments,
and the two best are shown here.
In both algorithms, the first step is to predict the $k$ cutoff.
If $k$ is greater than the threshold $T_k$, then proceed to
the \jass{} pipeline as in Algorithm~{\ref{alg-1}}, or make a
second query difficulty prediction as in Algorithm~{\ref{alg-2}}.
If \jass{} is used, a prediction for $\rho$ is made, but capped
at $\rhomax$, which allows us to achieve the desired performance
guarantees.
In our experiments, $\rhomax = 10$ million postings as this requires
less than $200$ms on our current hardware configuration.
The remaining queries are processed using \bmw{} with rank-safety.

\section{Experiments}
\label{sec-experiments}

We now look at the various predictions that are necessary to achieve
our performance requirements.
Our performance requirements for effectiveness are to achieve a target
$\MED$ that is low enough to result in no measurable effectiveness
difference for the target metric.
Our performance requirements for efficiency are no queries over
$200$ ms with a $99.99\%$ response time guarantee. That is, we can afford at
most $3$ over-budget queries for our entire query trace.

\begin{figure*}[t]
\centering
\includegraphics[width=\columnwidth]{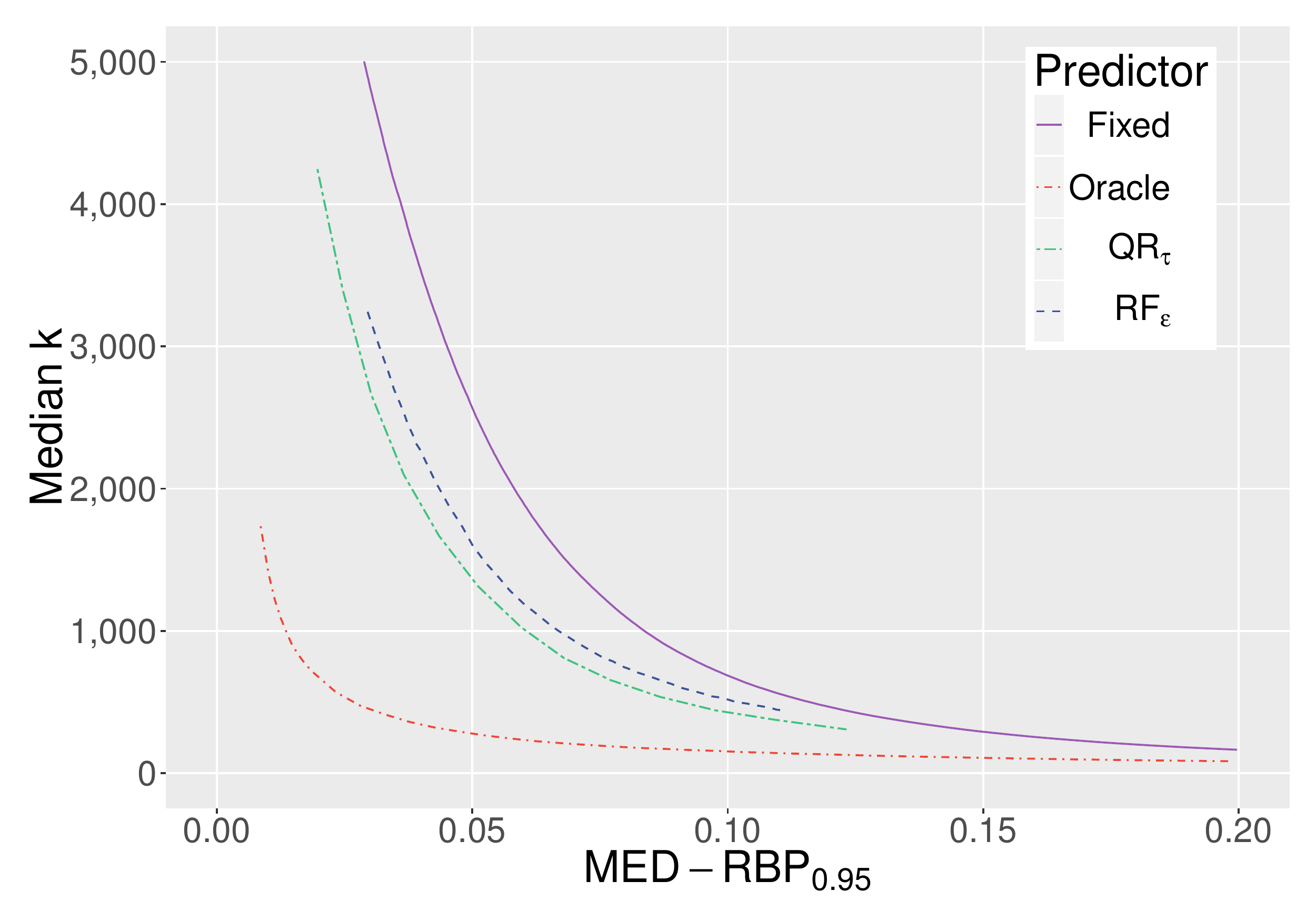}\hfill
\includegraphics[width=\columnwidth]{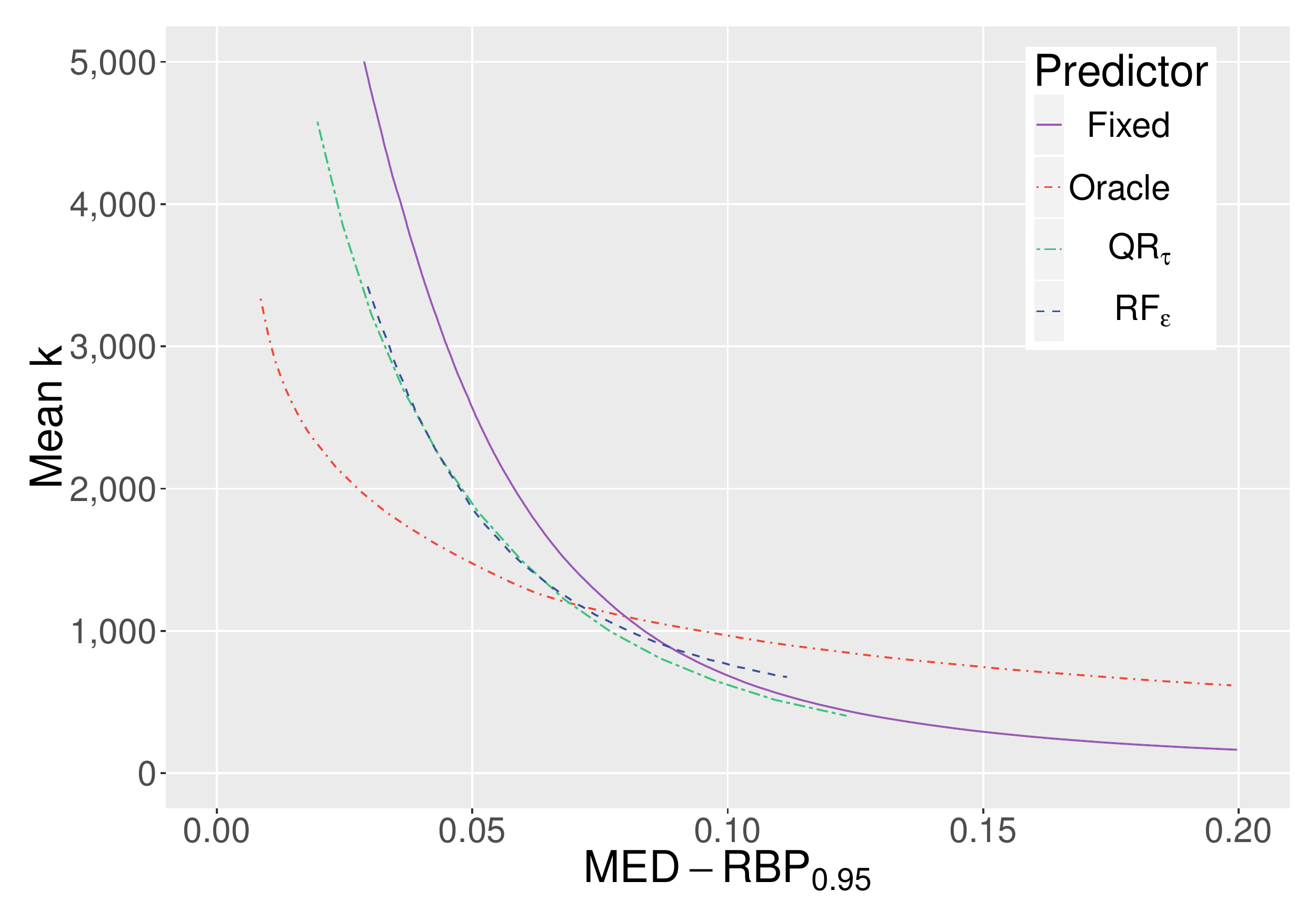}
\mycaption{
$\med{RBP}$ versus median $k$ (left) and mean $k$ (right) for all 
$\epsilon$ thresholds
between $0.001$ and $0.200$
when using a Random Forest regression, and for all $\tau$ values between
$0.10$ and $0.75$ with $\epsilon = 0.001$ for Quantile Regression,
in first stage retrieval for the $31{,}642$ queries from the
MQ2009 TREC Task. Note that the Quantile Regression clearly improves the
median $k$ (compared with Random Forests) without negatively affecting the 
mean $k$.
}\label{fig-krbp}
\vspace{-3mm}
\end{figure*}

\myparagraph{Predicting $k$}
First, we validate that our new approach to $k$ prediction using
quantile regression is effective. 
Using our newly devised regression technique, we can compare the
efficiency and effectiveness trade-offs between the size of the
candidate retrieval set $k$, and the expected effectiveness loss
$\med{RBP}$.
Figure~{\ref{fig-krbp}} shows the predictive power of the 
random forest ($\mbox{RF}_\varepsilon$) and quantile regression ($\mbox{QR}_\tau$) 
when compared to the oracle results for $\epsilon$ target between
$0.001$ and $0.10$, and to using a fixed cutoff for all queries.
Note that the graph on the left presents results as the median $k$
result in contrast to the right graph which shows the results for the
mean $k$ results as done in previous work.
Since the distribution of the true $k$ values is skewed for the queries
as shown in Figure~{\ref{fig-kdist}}, presenting the results using the
median more accurately captures the trade-offs.

\begin{figure}[t]
\centering
\includegraphics[width=\columnwidth]{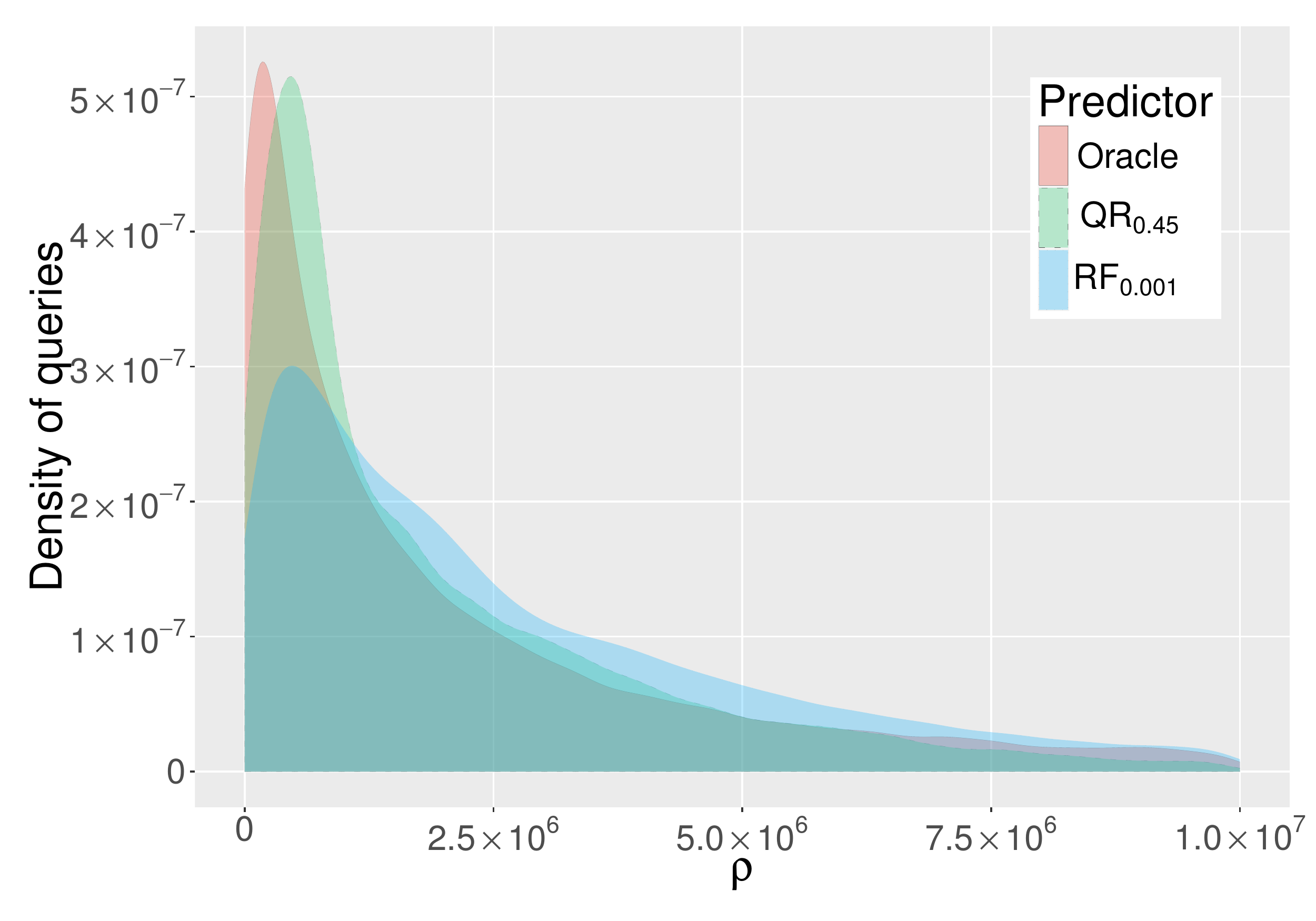}
\mycaption{A comparison of the distributions the actual $\rho$ vs the
predicted $\rho$ when using a Random Forest regression and a Quantile
Regression in first stage retrieval for the $31{,}642$ queries from
the MQ2009 TREC Task.
Note that the Random Forest uses a training value of $\epsilon =
0.001$, whereas the best-fit distribution for the Quantile Regression
was $\tau = 0.45$ for $\rho$.
}\label{fig-rhoden}
\end{figure}

\myparagraph{Predicting $\rho$}
Based on the lessons learned when attempting to build a robust
predictive framework for $k$, we now turn our attention to the
aggressiveness parameter $\rho$ in \jass{}.
Previous work has shown that using an exhaustive $\rho$ results in
effective top-$k$ retrieval, however, using a heuristic $\rho$ can
give similar effectiveness, yet much more efficient
retrieval~{\cite{lt15-ictir,jl16-ecir}.
The recommended heuristic value of $\rho$ is $10$\% of the size of
the collection~{\cite{lt15-ictir}}, which is around $5$ million for
the ClueWeb09B collection.
Figure~\ref{fig-rhoden}} shows the distribution of $\rho$ values
required to when targeting a MED-RBP$_{0.95}$ $< 0.001$, or
essentially, no measureable difference in the results lists between
exhaustive and aggressive \jass{} traversals.
Clearly, the majority of the distribution lies well to the lower side
of the $10$\% heuristic value.
This motivates us to predict $\rho$ on a query-by-query basis.
Again, we deploy both a Random Forest and a Gradient Quantile
Regression method as the distribution of $\rho$ is skewed.

\begin{figure}[t]
\centering
\includegraphics[width=\columnwidth]{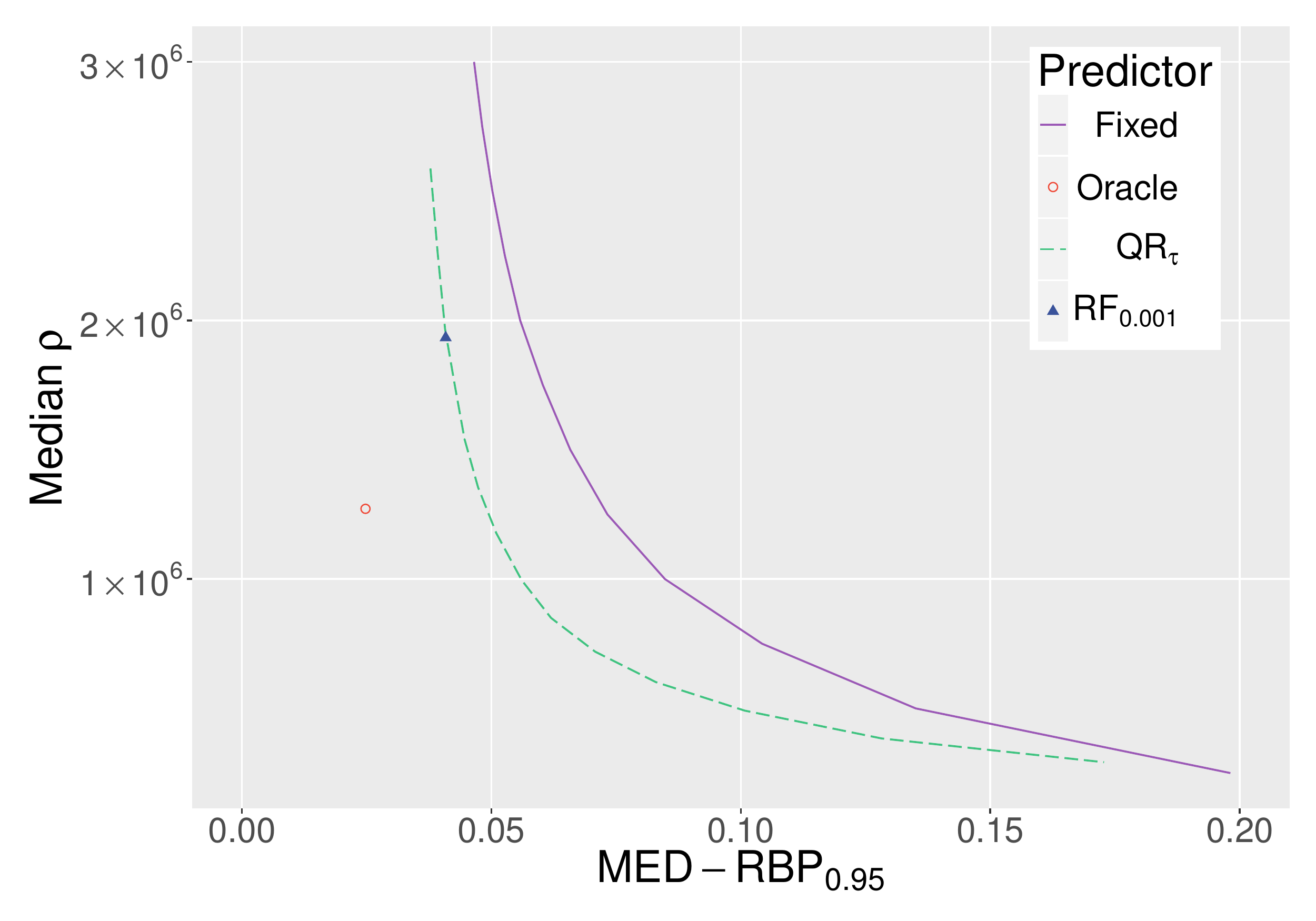}
\mycaption{
$\med{RBP}$ versus median $\rho$ for $\epsilon = 0.001$ when using the
RF regression, and for all $\tau$ values between $0.10$
and $0.75$ with $\epsilon = 0.001$ for QR, in first
stage retrieval for all $31{,}642$ queries.
Quantile Regression and Random Forests behave similarly with respect
to the median $\rho$, but QR is still preferred as the final
predicted $\rho$ distribution fits better with the idealized results
as shown in Figure~{\ref{fig-rhoden}}. 
}
\label{fig-rhopred}
\end{figure}

Figure~{\ref{fig-rhopred}} shows the median predicted $\rho$
values compared with the fixed and oracle. 
Both the QR and RF regression methods manage to improve on the fixed $\rho$
median.
Note that when measuring the MED-RBP$_{0.95}$ for this experiment
(and subsequently, training the value of $\rho$), the $k$ utilized
was the optimal value of $k$ from the previous experiment.
The reason for using this $k$ is that we must fix $k$, otherwise our
effectiveness scores may change as a result of $k$, not just $\rho$.
Indeed, this setting of $k$ also allows us to find the true optimal
MED-RBP$_{0.95}$ for \jass{}, denoted by the oracle point in
Figure~{\ref{fig-rhopred}}.

\begin{table}[t]
\begin{center}
\resizebox{0.48\textwidth}{!}{
\begin{tabular}{lccccccccc}
\toprule
\bf{System} & \bf{RMSE} & \bf{Precision} & \bf{Recall} & \bf{F} & \bf{M-Precision} & \bf{M-Recall} & \bf{M-F} & \bf{AUC}\\
\midrule
QR & \bf{0.76} & \bf{0.73} & 0.52 & \bf{0.62} & \bf{0.87} &\bf{0.76}&\bf{0.81}&\bf{0.98}\\
RF & 0.77 & 0.71 & \bf{0.54} & 0.61 & 0.84&0.76&0.80&0.97\\
LR & 0.84 & 0.73 & 0.49 & 0.58 & 0.85 &0.74&0.79&0.96\\
\bottomrule
\end{tabular}
}
\end{center}
\mycaption{Regression and tail query classification ($\tau=0.95$) performance for Quantile Regression, Random Forests and Linear Regression, best values bold (difference may be on the third decimal)}
\label{tbl-timing-pred}
\end{table}

\myparagraph{Predicting response time}
Given that our entire framework is built using query performance
prediction features, and we want to minimize tail-latency queries, we
explore the accuracy of query performance prediction within
the framework.

Table~\ref{tbl-timing-pred} shows the performance of three different
regression methods for regressed query times and for predicting
whether a query time will fall into the last percentile of the
distribution, i.e., if it will be a {\emph{tail-latency}} or not.
We replicate the previous set-up by using exactly the same features
as before and 10-fold cross validation.
We learn a regressor based on Random Forest, Gradient Quantile
Regression, and a Linear Regression, which was employed previously
by~\citet{mto12-sigir} for the same task, although with a smaller set
of features.

We report on regression performance using root mean squared error
(RMSE) and on a number of binary classification metrics for the tail-latency
prediction task.
To predict tail-latency queries for the $99$th percentile, we learn a
threshold in the training set by selecting the minimum running time
of all the queries in the $95$th percentile.
We report on the area under the curve (AUC), precision/recall/F measure for the
positive class (the query was a tail-latency query) and class-average (macro)
precision/Recall/F-measure.

Results show that our predictors are extremely effective for
regressing timings, with random forests and quantile regression
having a clear edge over linear regression, both in terms of raw
regression error (RMSE) and true positive classification.
QR has some advantage over RF given that the distribution of timings
is skewed (Figure~\ref{fig-tail}).
One discussion point is that we did not attempt to deploy any dynamic
features, such as those seen in the DDS prediction
framework~\cite{sk15-wsdm}. We leave this for future work.

\myparagraph{Putting it all together}
Here, we show that by combining all of our predictions into hybrid first-stage
retrieval systems, outlined in Algorithms~\ref{alg-1} and ~\ref{alg-2}, we can
achieve effectiveness equal to a fixed parameter system, while simultaneously
reducing the number of documents that must be passed on to the next stage of
the multi-stage retrieval system. Additionally, we show that we can use our
framework to mitigate tail-latency queries effectively.

\begin{figure}[t]
\centering
\includegraphics[width=\columnwidth]{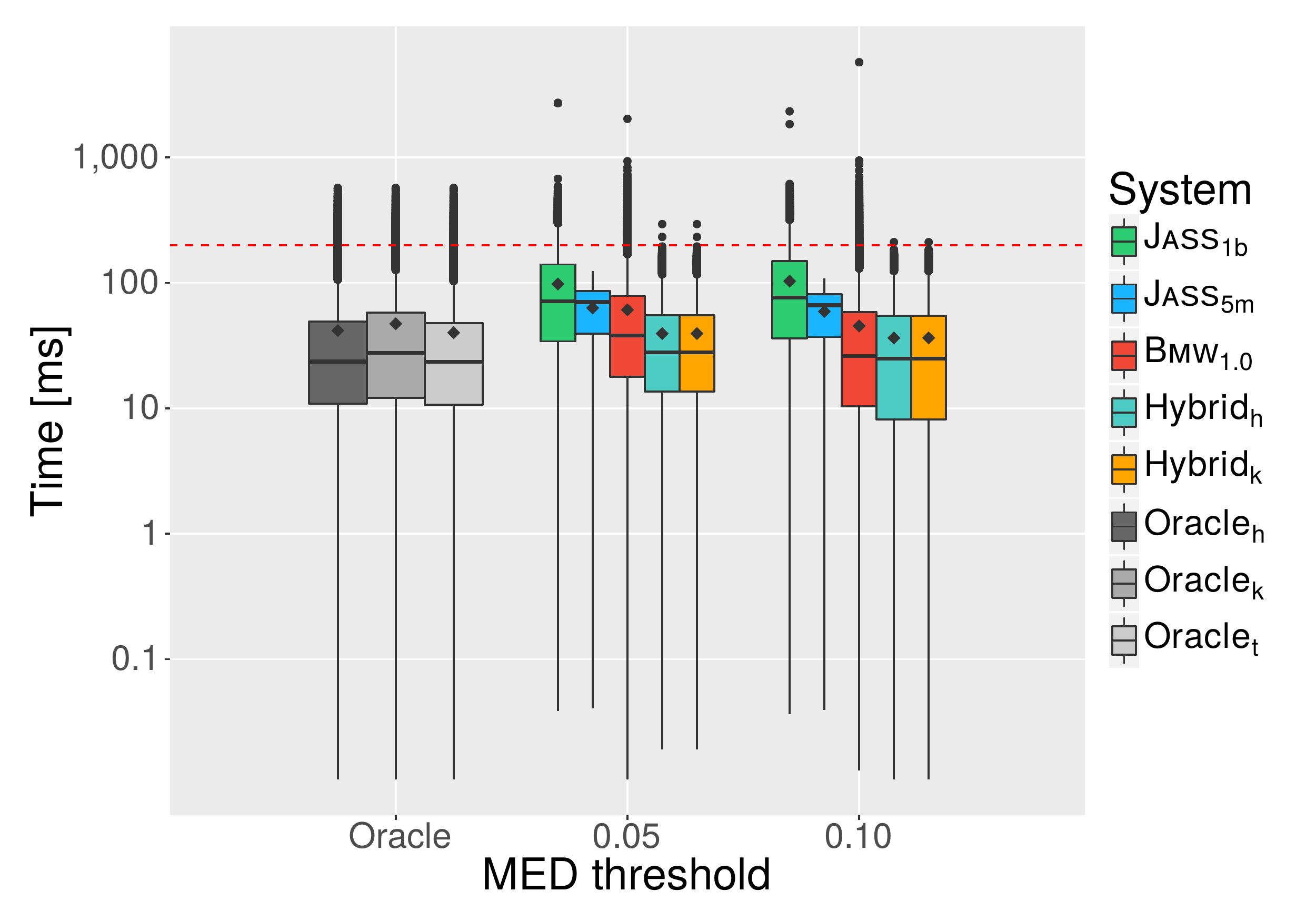}
\mycaption{The response time for each system for different bands of 
MED-RBP$_{0.95}$. Both Hybrid$_k$ and Hybrid$_h$ systems, which predict $k$ and $\rho$,
and $k$, $\rho$ and time respectively, show a clear improvement over the fixed
baselines. Although aggressive \jass{} has fewer tail latency queries, the $k$
required to attain the given MED values is larger than the exhaustive and hybrid
systems, which has implications in the efficiency of the late stage feature
extraction and re-ranking. The horizontal line denotes the $200$ms budget.}
\label{fig-moneybox}
\end{figure}



\begin{table}[t]
\begin{center}
\resizebox{0.48\textwidth}{!}{
	\begin{tabular}{lccccc}
\toprule
\multicolumn{6}{c}{{Oracles: MED-RBP$_{0.95} < 0.02$}} \\
\midrule
\bf{System} & \bf{Mean $k$} & \bf{Median $k$} & \bf{Mean time} & \bf{Median time} & \bf{\% queries $> 200$ ms}\\
\midrule
Oracle$_k$ & 3334 & 1735 & 47.2 & 27.6 & 3.1\\
Oracle$_t$ & 3334 & 1735 & \bf 40.0 & \bf 23.5 & \bf 2.2\\
Oracle$_h$ & 3334 & 1735 & 41.6 & 23.6 &  2.6\\
\midrule
\multicolumn{6}{c}{{MED-RBP$_{0.95} = 0.05$}} \\
\midrule
\bf{System} & \bf{Mean $k$} & \bf{Median $k$} & \bf{Mean time} & \bf{Median time} & \bf{\% queries $> 200$ ms}\\
\midrule
\bmw{}$_{1.0}$ & 2600 & 2600 & 60.9 & 38.1 & 4.4\\
\jass{}$_{1b}$ & 2600 & 2600 & 97.9 & 71.4 & 11.8\\
\jass{}$_{5m}$ & 3100 & 3100 & 63.1 & 70.3 & \bf 0\\
Hybrid$_k$ & \bf 2232 & \bf 1667 & \bf 40.9 & \bf 29.2 & 0.006\\
Hybrid$_h$ & \bf 2232 & \bf 1667 & \bf 40.9 & \bf 29.2 & 0.006\\
\midrule
\multicolumn{6}{c}{{MED-RBP$_{0.95} = 0.10$}}\\
\midrule
\bf{System} & \bf{Mean $k$} & \bf{Median $k$} & \bf{Mean time} & \bf{Median time} & \bf{\% queries $> 200$ ms}\\
\midrule
\bmw{}$_{1.0}$ & 800 & 800 &  45.3 & 26.1 & 2.4\\
\jass{}$_{1b}$ & 800 & 800 & 103.2 & 74.9 & 13.2\\
\jass{}$_{5m}$ & 900 & 900 & 59.2 & 65.7 & \bf 0\\
Hybrid$_k$ & \bf 648 & \bf 441 & \bf 36.4 & \bf 24.9 & 0.003\\
Hybrid$_h$ & \bf 648 & \bf 441 & \bf 36.4 & \bf 24.9 & 0.003\\
\bottomrule
\end{tabular}
}
\end{center}
\mycaption{Summary statistics for $k$, time and the \% of queries with response
times above $200$ ms. Each sub-table corresponds to a section of 
Figure~\ref{fig-moneybox}, and the best values are bold. Not only do the hybrid
systems require less documents in the first stage, they also run more efficiently
across the ISNs, and generally reduce tail latencies compared to fixed systems. In particular, the hybrid methods both have only 1 query $> 200$ms in the
MED $0.10$ case, and 2 queries $> 200$ms in the MED $0.05$ case.}
\label{tbl-money}
\end{table}

Figure~{\ref{fig-moneybox}} shows the 
performance for $2$ different MED-RBP$_{0.95}$ cut-offs: $0.05$ and $0.10$.
We also show the performance of the oracle selectors, which all had
MED-RBP$_{0.95}$ scores below $0.02$.
As before, \jass{}$_{1b}$, \jass{}$_{5m}$ and
\bmw{}$_{1.0}$ refer to using a fixed $k$ -- the $k$ was selected
such that the mean MED value was equivalent to the target.
We also report the results of the two hybrid systems based on
Algorithm~\ref{alg-1} (Hybrid$_k$) and Algorithm~\ref{alg-2} (Hybrid$_h$), 
which use quantile regression for their predictions.
Additionally, Table~{\ref{tbl-money}} shows the average and median $k$, as
well as the time
characteristics for the systems presented in Figure~{\ref{fig-moneybox}}. 

Our results show that our hybrid systems both outperform the equivalent
fixed \bmw{} or \jass{} traversals for the given MED targets. For example,
with a target of MED-RBP$_{0.95} = 0.05$, our hybrid systems can achieve
a mean and median query response time $20$ ms and $8.9$ ms below the best fixed 
system, respectively. The hybrid
systems return, on average, $368$ less candidate documents to the next stage
of the retrieval architecture, resulting in further efficiency gains along the
cascade without loss in effectiveness. Finally, our hybrid systems managed
to each have only $2$ queries that ran longer than our target efficiency 
of $200$ ms, with run times of $232.4$ ms and $294.1$ ms respectively. 
Similar outcomes are observed when the MED target is relaxed to $0.10$.
Although the \jass{}$_{5m}$ fixed system
outperforms our hybrids in reducing tail latencies, it must retrieve a larger
number of documents to achieve the same effectiveness target, which has negative
implications on the efficiency of the following stages.
We note that we do not consider the time required to make our predictions. Recent
work using similar models show a prediction overhead of $< 0.75$ ms per 
prediction~\cite{jk+14-sigir}. So, in the worst case, we are likely to only add 
$2-3$ms per query.


\myparagraph{Validating Robustness}
As a final test of robustness, we run both of our hybrid systems
across
the $50$ (unseen) TREC 2009 Web Track queries. These queries were held out
from the train and test procedures reported in earlier sections. Since
these queries have judgements to depth $12$, we report NDCG@$10$, ERR@$10$
and RBP$_{p=0.80}$~\cite{lmc16-irj}.
For the hybrid systems, we used the same prediction configuration that was used
in the MED-RBP$_{0.95} = 0.05$ task from Figure~{\ref{fig-moneybox}} and 
Table~{\ref{tbl-money}}.

Table~\ref{tbl-validate} shows the effectiveness measurements. Clearly,
our hybrid systems have a small loss in effectiveness compared to the
ideal end-stage run. In order to test whether the {\tt uog-ideal} run was
significantly better than our hybrid runs, we ran the 
two one-sided test~\cite{s87-tost} of equivalence (TOST). For each TOST
test, we set the $\epsilon$ parameter as $\epsilon = 0.1 \cdot \mu$, where $\mu$ is
the mean effectiveness score of the ideal run for the desired metric.
We found that the  ideal system was not statistically
significantly different than our hybrid systems, with $p < 0.05$.

\begin{table}[t]
\begin{center}
\resizebox{0.48\textwidth}{!}{
\begin{tabular}{l c c c}
\toprule
\bf System & \bf NDCG@$10$ & \bf ERR@$10$ & \bf RBP $p=0.80$ \\
\midrule
{\tt uog-ideal} & 0.3578 & 0.4346 & 0.4357 (0.1366)\\
Hybrid$_{k}$ & 0.3464 & 0.4174 & 0.4231 (0.1523)\\
Hybrid$_{h}$ & 0.3464 & 0.4174 & 0.4231 (0.1523)\\
\jass{}$_{5m}$ & 0.3554 & 0.4354 & 0.4297 (0.1517)\\
\bottomrule
\end{tabular}
}
\end{center}
\mycaption{Effectiveness measurements taken across the held-out query set.
No statistical
significance was measured between the hybrid systems with respect to the
ideal system, using the two one-sided test with $p < 0.05$.
}
\label{tbl-validate}
\end{table}

\section{Conclusion}
\label{sec-conclusion}
We presented and validated a unified framework to predict a
wide range of performance-sensitive parameters for early-stage candidate retrieval
systems using $\MED$ and reference lists as guides for training (RQ1). 
Preliminary experiments show that the \daat{} \bmw{} approach is efficient
but suffers from the occasional tail query, which the \saat{} \jass{} algorithm does
not. Hybrid systems based on this
framework were shown to minimize
effectiveness loss while also minimizing query-latency across all stages of a
multi-stage search architecture. Given a fixed budget of $200$ms for a
first-stage response time, we can achieve this budget $99.99\%$ of the
time with the hybrid systems, across an index of $50$ million documents and
a trace of over $30{,}000$ queries, thus answering RQ2 in the affirmative.
In particular, we find that using quantile regression (GBRT) for predicting
$k$, $\rho$ and response time allows us to minimize the late stage effectiveness
loss while simultaneously minimizing the size of the initial candidate set, thus
answering RQ3.


\begin{small}
\setlength{\bibsep}{2.0pt}
\bibliographystyle{ACM-Reference-Format}
\bibliography{strings-shrt,local}
\end{small}

\end{document}